\documentclass[aps,prc,twocolumn,superscriptaddress]{revtex4-1}

\usepackage{amsmath,amssymb,graphicx}
\usepackage{url}
\usepackage[colorlinks,urlcolor=blue]{hyperref}

\begin{document}

\title{Energy-energy correlators inside single inclusive jets in heavy-ion collisions with CoLBT-hydro model}

\author{Zhong Yang}
\affiliation{Department of Physics and Astronomy, Vanderbilt University, Nashville, TN, 37235, USA}

\author{Raghav Kunnawalkam Elayavalli}
\affiliation{Department of Physics and Astronomy, Vanderbilt University, Nashville, TN, 37235, USA}

\author{Xin-Nian Wang}
\affiliation{Key Laboratory of Quark and Lepton Physics (MOE) \& Institute of Particle Physics, Central China Normal University, Wuhan 430079, China}

\begin{abstract} 
The energy-energy correlator (EEC) inside jets is a sensitive observable for studying jet modification in the quark-gluon plasma (QGP). However, its interpretation in heavy-ion collisions remains challenging, requiring a consistent understanding of jet evolution across multiple dynamical scales together with a proper treatment of the background subtraction. In this work, we employ an updated CoLBT-hydro framework in which a medium scale $Q_M$ = 2.0 GeV is introduced to separate the vacuum and in-medium stages of the parton shower, enabling a more self-consistent treatment of jet evolution. Using a theoretical background subtraction within the model, the resulting simulation reproduces the recent CMS measurement of the in-jet EEC, and through a decomposition of different contributions, highlights the impact of medium modification on the observable. To further validate the experimental procedure, we also implement the CMS mixed-event background-subtraction method directly in the simulation and find the results are consistent with those obtained with the theoretical background subtraction. Using $p_T$-ranked jets in each event, we further investigate the dependence of medium modification on the in-medium path length, reflected in the different EECs of leading and sub-leading jets. Finally, we explore the dependence of the leading-jet EEC on the dijet rapidity gap as a signal of the jet-induced diffusion wake.


\end{abstract}

\maketitle

\section{Introduction}
High-energy heavy-ion collisions produce a strongly coupled QGP. Jets originating from hard parton scattering processes traverse the medium and lose a substantial fraction of their energy, a phenomenon known as jet quenching~\cite{Bjorken:1982tu, Thoma:1990fm, Braaten:1991we, Gyulassy:1993hr, Baier:1996kr, Zakharov:1996fv, Gyulassy:1999zd, Wiedemann:2000za, Wang:2001ifa, Arnold:2002ja, Djordjevic:2006tw, Qin:2007rn, Wang:2025lct}, which is observed at both RHIC~\cite{PHENIX:2001hpc, STAR:2002ggv, Gyulassy:2004zy, Wang:2004dn} and the LHC~\cite{ALICE:2010yje, ATLAS:2010isq, Majumder:2010qh, CMS:2011iwn, Qin:2015srf}, providing one of the compelling evidence for the formation of QGP in high-energy heavy-ion collisions.

Jet substructure has emerged as a central topic in jet physics~\cite{Apolinario:2024equ}.
They offer a powerful handle on jet evolution in the QGP and deepen our understanding of quantum chromodynamics (QCD) under extreme conditions. As a key jet-substructure observable, the EEC was originally introduced in $e^+e^-$ annihilation~\cite{Basham:1977iq, Basham:1978bw, Basham:1978zq}. 
After extension to heavy-ion collisions~\cite{Komiske:2022enw, Moult:2025nhu}, it has attracted considerable interest~\cite{Andres:2022ovj, Yang:2023dwc, Barata:2023bhh, Andres:2023xwr, Andres:2023ymw, Barata:2024wsu, Fu:2024pic, Andres:2024xvk, Xing:2024yrb, Singh:2024vwb, Bossi:2024qho, Andres:2024hdd, Barata:2024ukm, Ke:2025ibt, Yang:2025bub, Barata:2025uxp, Barata:2025jhd, Budhraja:2025ulx, Barata:2025fzd, Liu:2025ufp, Apolinario:2025vtx, Andres:2025yls}, and measurements have been performed in different collision systems and at different colliding energies~\cite{ALICE:2024dfl, CMS:2024mlf, STAR:2025jut, ALICE:2025igw, CMS:2025ydi, CMS:2025jam, Nambrath:2025ttz}. The EEC is particularly powerful because its angular dependence can separate perturbative and nonperturbative regimes~\cite{Liu:2024lxy, Barata:2024apg, Lee:2024jnt, Jaarsma:2025tck, Guo:2025qnz, Guo:2025zwb, Chen:2025rjc, Electron-PositronAlliance:2025wzh, Alipour-fard:2025dvp, Budhraja:2026pyi, Duan:2026icj}, thereby probing jet evolution across different dynamical scales. 
Theoretically, it has been shown to be sensitive to a broad range of physics, including cold nuclear matter effects in small systems~\cite{Ke:2025ibt, Yang:2025bub, Barata:2024wsu, Fu:2024pic, Andres:2024xvk}, the onset of color coherence~\cite{Andres:2022ovj, Andres:2023xwr}, heavy-quark mass effects~\cite{Andres:2023ymw, Barata:2025uxp, Xing:2024yrb}, jet quenching and medium response~\cite{Barata:2023bhh, Bossi:2024qho, Andres:2024hdd, Barata:2024ukm, Barata:2025fzd, Liu:2025ufp, Yang:2023dwc, Apolinario:2025vtx, Andres:2025yls, Barata:2026pgh}.
These studies have established the EEC as a valuable tool for constraining the in-medium evolution of jets and the properties of the QGP.

Since the jet dynamics probed by EEC spans from the collinear to the wide-angle regime, a quantitative description of EEC in high-energy heavy-ion collisions over such a wide range of angular scales remains a challenge, requiring a consistent modeling of the multi-scale evolution of jets in the medium, including how parton splittings are modified and how the jet-induced medium response develops. In experimental and phenomenological studies, an additional complication arises from the large QGP background, whose contribution to the measured EEC is substantial and must be carefully subtracted before a meaningful comparison between theory and experiment can be made~\cite{CMS:2025ydi}.

In this work, we employ an updated CoLBT-hydro framework to calculate the EEC inside single-inclusive jets and compare our results with the recent CMS measurements. To properly account for the multi-stage evolution of jet showers, the framework allows medium-modified partons to continue their vacuum evolution after exiting the QGP. To enable a more realistic comparison with data, we further implement the mixed-event background-subtraction procedure used by CMS and compare it with the idealized theoretical subtraction. We then classify jets according to their $p_T$ rank within each event, revealing the sensitivity of the EEC to the effective in-medium path length since leading and sub-leading jets traverse different paths in the QGP medium. Finally, we investigate the dependence of the leading-jet EEC on the dijet rapidity gap and its connection to the jet-induced diffusion wake.


\section{EEC inside single inclusive jets}
The CoLBT-hydro model combines the Linear Boltzmann Transport (LBT) model~\cite{Wang:2013cia, He:2015pra, Cao:2016gvr, Cao:2017hhk, Luo:2023nsi, Xing:2019xae} with event-by-event (3+1)D CCNU-LBNL viscous (CLVisc) hydrodynamics model~\cite{Pang:2012he, Pang:2014ipa, Pang:2018zzo, Wu:2021fjf}, providing a concurrent description of the evolution of hard partons and soft hydro response in the QGP. More details can be found in Refs.~\cite{Chen:2017zte, Chen:2020tbl, Zhao:2021vmu}. In the original CoLBT-hydro implementation, a parton shower in PYTHIA8~\cite{Sjostrand:2007gs, Sjostrand:2014zea, Bierlich:2022pfr} is evolved in vacuum down to the hadronization scale $Q_0 = 0.5$~GeV before it starts interacting with the medium. As a consequence, the vacuum shower is fully completed prior to any medium interaction, which is inconsistent with the physical picture in which medium-modified partons experience momentum broadening and can continue to radiate in vacuum after exiting the QGP~\cite{Andres:2024egc}. To address this, we introduce a medium scale $Q_M = 2.0$~GeV, following Refs.~\cite{Ke:2020clc, Dang:2026ezw}, which separates the vacuum and in-medium stages of the shower evolution. The jet shower in PYTHIA8 is first evolved in vacuum down to $Q_M$ and then propagates through the medium according to CoLBT-hydro. After exiting the QGP,  shower partons are handed back to PYTHIA8 for further vacuum evolution down to $Q_0$, followed by hadronization.

We use the CoLBT-hydro model to simulate the transport of jets through the QGP in $0$--$10\%$ Pb+Pb collisions at $\sqrt{s_{\mathrm{NN}}}=5.02$~TeV. Jets are reconstructed with the anti-$k_T$ algorithm in the E-scheme using FASTJET~\cite{Cacciari:2011ma}. Their constituents are then reclustered with the winner-take-all (WTA) scheme~\cite{Bertolini:2013iqa} to define the jet axis used in the calculation of EEC. Particles within $R<0.4$ of this axis, where $R=\sqrt{(\eta_{\mathrm{jet}}-\eta_h)^2+(\phi_{\mathrm{jet}}-\phi_h)^2}$, are selected to construct the EEC distribution:
\begin{equation}
    \mathrm{EEC}(R_L)=\frac{1}{W_{\mathrm{pairs}}}\frac{1}{\delta r}
    \sum_{\mathrm{jets}}
    \sum_{\substack{i,j\in\mathrm{jet} \\ r_{ij}\in\mathrm{bin}(R_L)}}
    \left(p_{T,i}p_{T,j}\right)^{n},
\end{equation}
where $p_{T,i}$ and $p_{T,j}$ are the transverse momenta of charged hadrons $i$ and $j$, respectively, and $\delta r$ denotes the bin width in the angular distance
\begin{equation*}
    r_{ij}=\sqrt{(\eta_{h,i}-\eta_{h,j})^2+(\phi_{h,i}-\phi_{h,j})^2}.
\end{equation*}
The normalization factor $W_{\mathrm{pairs}}$ is defined as the sum of $(p_{T,i}p_{T,j})^{n}$ over all selected hadron pairs. In this work, we take $n=1$, which corresponds to the nominal EEC used in the CMS analysis~\cite{CMS:2025ydi}.

\begin{figure}[h!]
\centering
    \includegraphics[width=0.45\textwidth]{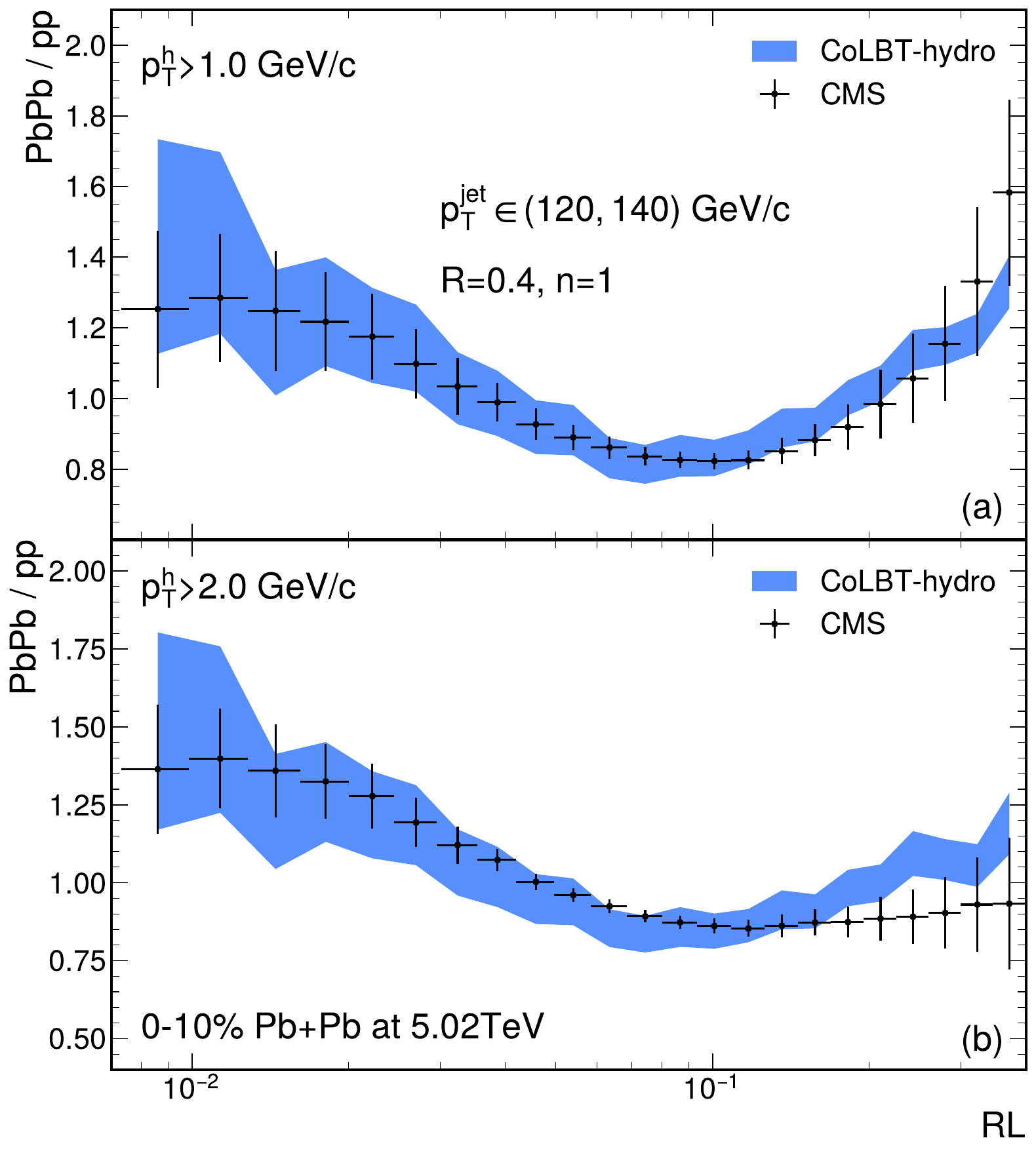}
    \caption{Ratio of the EEC in single-inclusive jets in $0$--$10\%$ Pb+Pb collisions to that in p+p collisions at $\sqrt{s_{\mathrm{NN}}}=5.02$~TeV for $p_T^{\mathrm{jet}}\in(120,140)~\mathrm{GeV}/c$ (a) with $p_T^{h}>1.0~\mathrm{GeV}/c$ and (b)  $p_T^{h}>2.0~\mathrm{GeV}/c$.}
    \label{EEC_Co}
\end{figure}
In the CoLBT-hydro framework, final-state hadrons receive contributions from both hard  partons in the LBT component and soft hadrons from the hydro response induced by jet through energy-momentum deposition in the QGP. To isolate the contribution to the EEC from  jet-induced hydro response, we perform two CoLBT-hydro simulations with identical hydrodynamic initial conditions, one with the jet and one without. The difference between the two CLVisc hydrodynamic outputs defines the jet-induced hydro response. This procedure provides an idealized theoretical background subtraction in the model calculation. From the resulting momentum spectrum, particles are sampled accordingly. Since the hydro response contains both positive and negative contributions, associated with the wake front and the diffusion wake~\cite{Gubser:2007ni, Betz:2008ka, Chen:2021gkj, Yang:2022nei, ATLAS:2024prm, CMS:2025dua, Yang:2025dqu, CMS:2026mur, Yang:2025lii, Yang:2025xni} of the jet-induced Mach-cone-like medium excitation~\cite{Casalderrey-Solana:2004fdk, Stoecker:2004qu, Ruppert:2005uz, Gubser:2007ga, Neufeld:2008fi, Qin:2009uh, Li:2010ts, Bouras:2012mh, Ayala:2016pvm, Yan:2017rku, Casalderrey-Solana:2020rsj}, the sampled particles are classified into positive and negative groups. The EEC then receives contributions from positive-positive, negative-negative, and positive-negative pairs, denoted as $\mathrm{EEC}_{pp}$, $\mathrm{EEC}_{nn}$, and $\mathrm{EEC}_{pn}$, respectively. The final EEC distribution is given by
\begin{equation}
    \mathrm{EEC} = \mathrm{EEC}_{pp} + \mathrm{EEC}_{nn} - \mathrm{EEC}_{pn}.
\end{equation}
Fig.~\ref{EEC_Co} shows the ratio of the EEC inside single inclusive jets in $0$--$10\%$ Pb+Pb collisions to that in p+p collisions at $\sqrt{s_{\mathrm{NN}}}=5.02$~TeV. The results are shown for jet transverse momentum $p_T^{\mathrm{jet}}\in(120,140)~\mathrm{GeV}/c$, jet cone-size $R=0.4$, and pseudorapidity  $|\eta_{\mathrm{jet}}|<1.6$, and are compared with the CMS data. We find that the updated  CoLBT-hydro calculation provides a good description of the CMS measurement for both transverse-momentum cuts $p_T^{h}>1~\mathrm{GeV}/c$ [Fig.~\ref{EEC_Co}(a)] and $p_T^{h}>2~\mathrm{GeV}/c$ [Fig.~\ref{EEC_Co}(b)] for charged hadrons. For the same final  $p_T^{\mathrm{jet}}$ selection, the initial  $p_T$ scale of the jet production in Pb+Pb collisions is larger than that in p+p collisions due to jet energy loss. The enhancement at small angles arises mainly from this  $p_T^{\mathrm{jet}}$ selection bias between Pb+Pb and p+p collisions, together with the change in the quark/gluon jet fraction. In contrast, the enhancement at large angles is dominated by the jet-induced medium response. As the charged-hadron $p_T$ cut is increased, the large-angle enhancement becomes significantly suppressed.

\begin{figure}[h!]
\centering
    \includegraphics[width=0.45\textwidth]{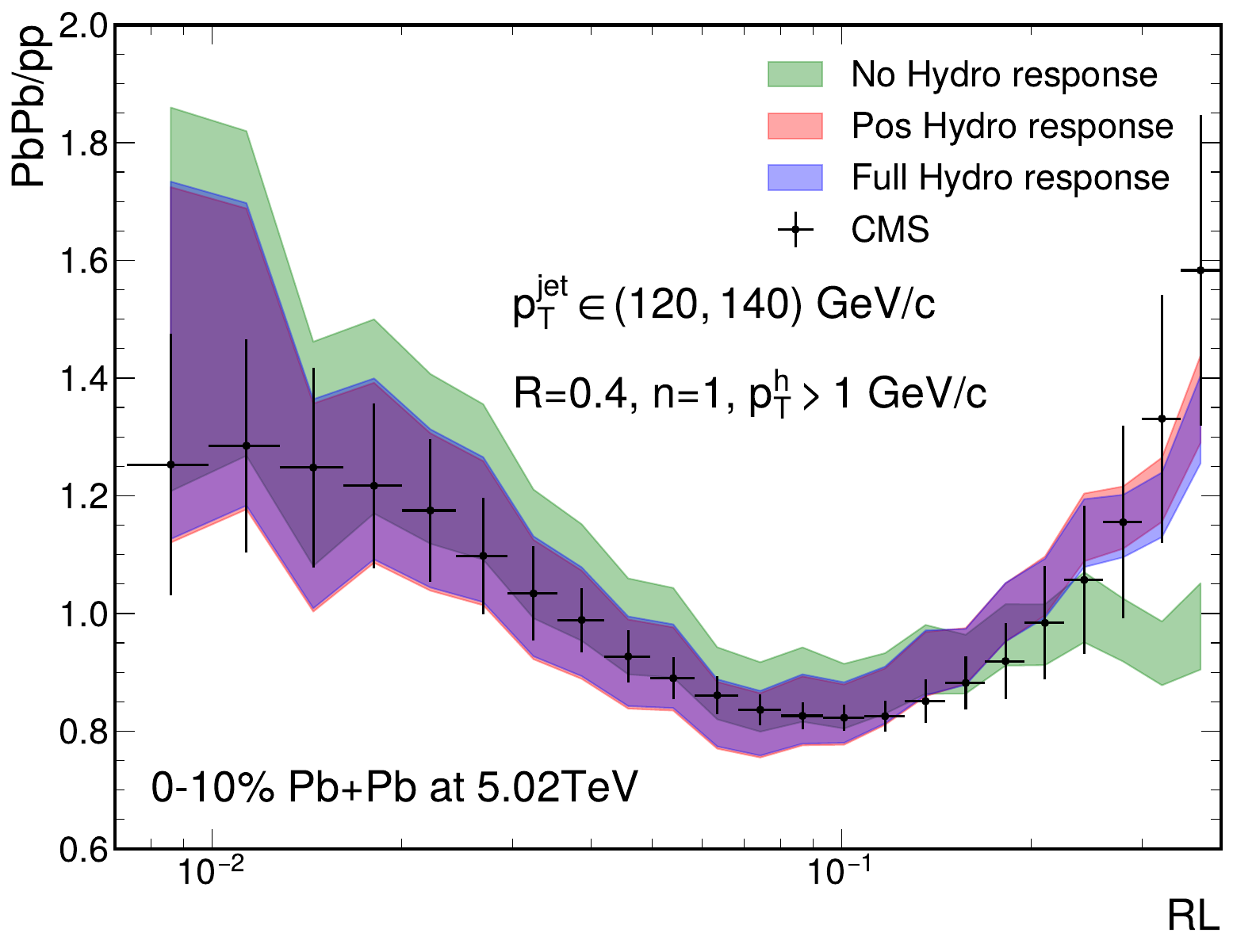}
    \caption{EEC distributions for three cases: including the full hydro response (blue band), including only the positive hydro-response contribution (red band), and excluding the hydro response (green band).}
    \label{EEC_hr}
\end{figure}

To further understand this behavior, we compute the EEC in three scenarios, including all hydro-response particles (blue band), including only positive particles (red band), and excluding the hydro response (green band), as shown in Fig.~\ref{EEC_hr}.
Comparing the blue and red bands, we find that the contribution from negative particles to the in-jet EEC is negligible. This behavior reflects the fact that, within the jet cone, the hydro response is dominated by the wake front of the Mach cone, while the diffusion wake is largely masked and contributes only weakly. In contrast, comparison of the blue and green bands shows that the hydro response significantly affects the EEC at large angles, leading to a clear enhancement in this region. This is consistent with previous studies of medium modifications to the EEC~\cite{Yang:2023dwc, Andres:2022ovj}.\\

\section{EEC with mixed-event background subtraction}
In experimental measurements of the in-jet EEC, background particles from the QGP can have a non-negligible impact as in the model calculations. However, the final signal and background particles cannot be distinguished in experiment, rendering the idealized theoretical subtraction used in the model calculations impossible. Since particles inside the jet cone contain both signal and background components, denoted by $S$ and $B$, respectively, the measured EEC contains three contributions: signal--signal ($SS$), signal--background ($SB$), and background--background ($BB$). The latter two contributions must be subtracted in order to obtain the true signal.  To address this, CMS employs a mixed-event background-subtraction method in its EEC measurement. Two mixed events, denoted by $M_1$ and $M_2$, are selected to have similar background particles to those of the jet event, and are assumed to approximate the background component $B$ in the jet event. The true signal is extracted via the combination
\begin{equation}
\begin{aligned}
    \mathrm{EEC}_{\mathrm{sig}} ={}& \langle (S{+}B)(S{+}B)\rangle - \langle (S{+}B)\,M_1\rangle \\
                                  &+ \langle M_1 M_2\rangle - \langle M_1 M_1\rangle.
\end{aligned}
\end{equation}
Here, the first term contains the desired signal $SS$ along with the $SB$ and $BB$ contaminations. The second term subtracts the $SB$ correlation using $M_1$ as a proxy for $B$, but simultaneously introduces a spurious $BM_1$ term. The third term, built from two independent mixed events, cancels this $BM_1$ residual, and the fourth term removes the remaining $BB$ contamination from the first term, leaving only the signal $SS$.

\begin{figure}[h!]
\centering
    \includegraphics[width=0.45\textwidth]{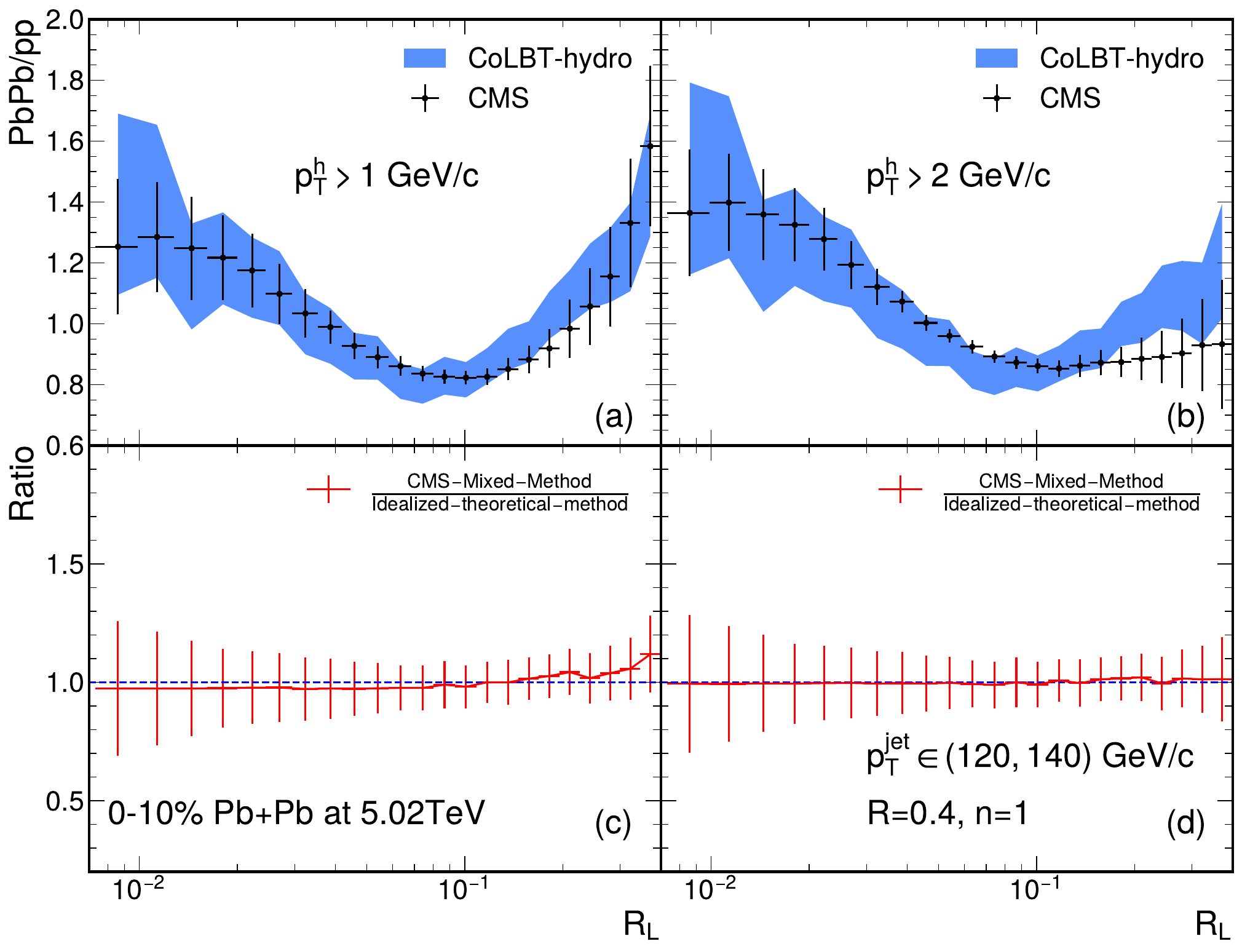}
    \caption{Ratio of the EEC in Pb+Pb to that in p+p collisions obtained with the mixed-event background-subtraction procedure (a) for $p_T^{h}>1~\mathrm{GeV}/c$ and (b) $p_T^{h}>2~\mathrm{GeV}/c$, respectively. Panels (c) and (d) show the ratio of these results to those obtained with the theoretical background-subtraction method in Fig.~\ref{EEC_Co}. 
    }
    \label{Method_compare}
\end{figure}
We can implement the same mixed-event subtraction procedure in our model simulation. Following the experimental method, for each triggered dijet event, the mixed events $M_{1}$ and $M_{2}$ are selected from a pool of events with charged-particle multiplicity matching that of the triggered event within 1.5$\%$, so that the mixed events are statistically compatible with the background of the signal event. The EEC is then recalculated exactly as in the experimental analysis. Fig.~\ref{Method_compare}(a) and (b) show the resulting Pb+Pb/p+p ratios obtained with the mixed-event subtraction for $p_{T}^{h}>1~\mathrm{GeV}/c$ and $p_{T}^{h}>2~\mathrm{GeV}/c$, respectively. Panels~(c) and~(d) show the ratios of these results to those of Fig.~\ref{EEC_Co}, where the idealized theoretical background subtraction is employed. Over most of the $R_{L}$ range, the two procedures yield nearly identical results, with a small residual deviation appearing only at large angles. This large-angle deviation reflects the intrinsic limitation of multiplicity-based mixing. The selected $M_{1}$ and $M_{2}$ events provide only an approximate representation of the background in the triggered event, leading to a small residual bias in the background estimate. As shown in the CMS analysis, the correlator at large $R_{L}$ is strongly affected by pairs involving background particles, making this region particularly sensitive to biases in the background estimate. Imposing the stricter $p_{T}^{h}>2~\mathrm{GeV}/c$ cut suppresses the soft-particle yield and visibly reduces this discrepancy, as seen in panel~(d). These remaining differences lie well within current experimental uncertainties and demonstrate the robustness of the CMS mixed-event subtraction. Our study here therefore provides independent support for this experimental procedure and strengthens confidence in its application to future energy-correlator measurements in heavy-ion collisions.\\

\section{EEC of leading and sub-leading jets}
\begin{figure}[h!]
\centering
    \includegraphics[width=0.42\textwidth]{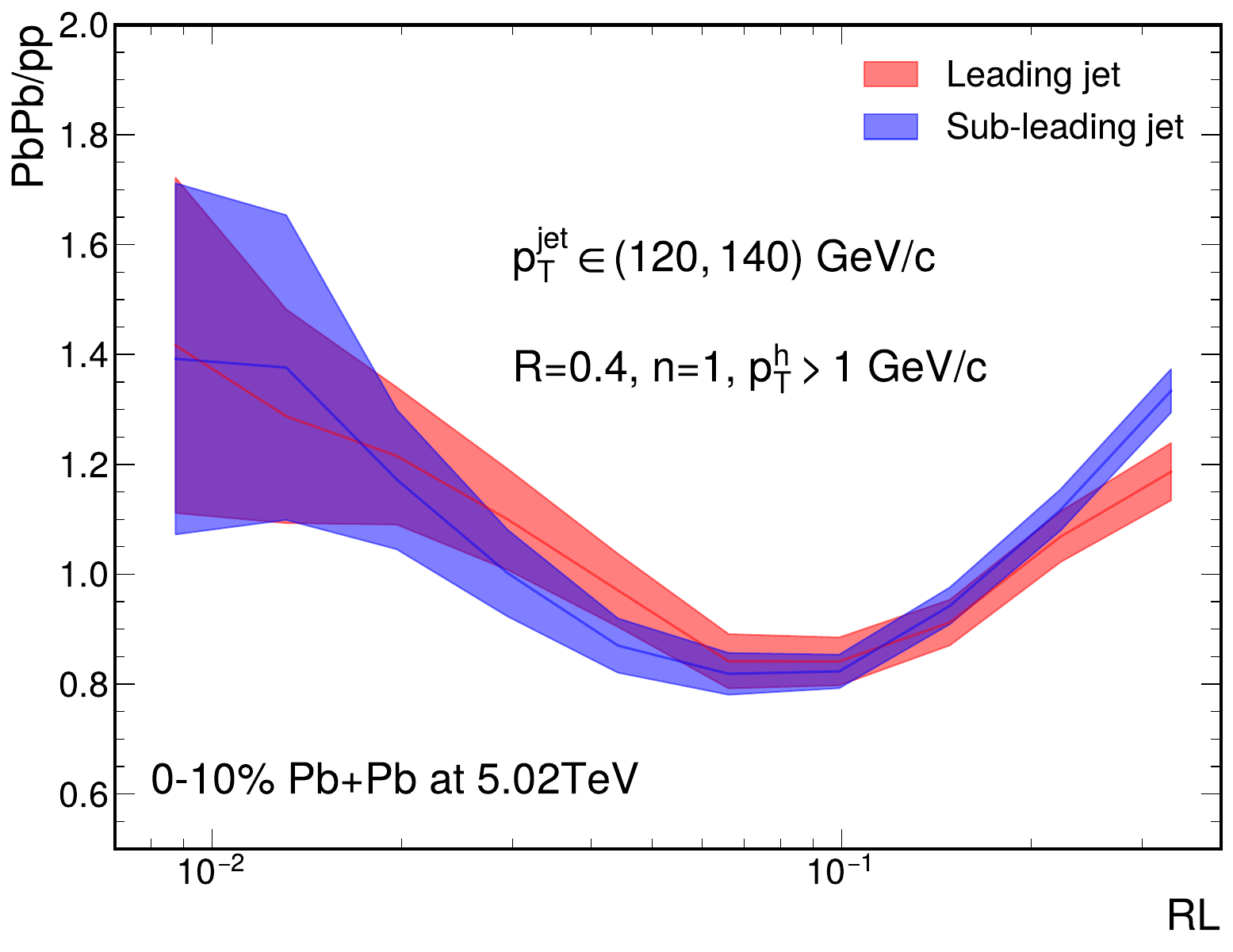}
    \caption{Ratio of the EEC inside leading (red) and sub-leading (blue) jets in $0$--$10\%$ central Pb+Pb collisions to that in p+p collisions at $\sqrt{s_{\mathrm{NN}}}=5.02$~TeV, for $p_T^{\mathrm{jet}}\in(120,140)~\mathrm{GeV}/c$, $R=0.4$, and $p_T^{h}>1~\mathrm{GeV}/c$. }
    \label{EEC_LS}
\end{figure}

Single inclusive jets are biased toward jet production near the surface of the QGP fireball and therefore probe a relatively short path length through the medium. To further explore the path-length dependence of the EEC, we rank jets in each event by their $p_T^{\mathrm{jet}}$ and divide them into leading (highest $p_T^{\mathrm{jet}}$) and sub-leading (second-highest) jets.  Since leading and sub-leading jets are dominated by back-to-back dijet pairs, the leading jet escapes the QGP region with a short path length due to trigger bias while the sub-leading jet traverses the whole length of the QGP. The corresponding EEC distributions are then computed and compared in Fig.~\ref{EEC_LS}.
The ratio of EEC in Pb+Pb over p+p collisions for sub-leading jets exhibits a larger wide-angle enhancement than that for leading jets, reflecting their longer in-medium path length and consequently stronger interaction with the QGP. The longer propagation length leads to increased energy loss and a more pronounced redistribution of energy to large angles through medium-induced radiation and jet-induced medium response. At small $R_{L}$, two competing effects shape the sub-leading jet ratio. On the one hand, the larger energy loss induces a stronger jet $p_{T}$ selection bias, which shifts the EEC distribution toward smaller angles and thereby enhances EEC in the small-$R_{L}$ region. On the other hand, the small-$R_{L}$ region is dominated by hard-particle correlations, which are suppressed by jet energy loss, reducing the EEC distribution in this region. The two effects compete, resulting in no significant difference between the modification of EEC of leading and sub-leading jet at small~$R_{L}$. \\

\section{Mapping the jet-induced diffusion wake with the EEC}
The medium modification of the EEC also provides a route to identifying the diffusion-wake signal, an essential component of the jet-induced medium response that carries direct information on the transport properties of the QGP. In dijet events, the diffusion wake of one jet depletes energy within the cone of its partner, with the strongest effect  at small dijet rapidity gap~\cite{Pablos:2019ngg, Yang:2025dqu}, thereby amplifying the partner’s effective energy loss and jet $p_T$ selection bias~\cite{ALICE:2024dfl, Andres:2024hdd}. Since the ratio of the EEC in Pb+Pb to p+p collisions is strongly shaped by this bias, comparing the EEC inside leading jets at different dijet rapidity gaps provides a novel method to search for the diffusion-wake signal.

\begin{figure}[h!]
\centering
    \includegraphics[width=0.48\textwidth]{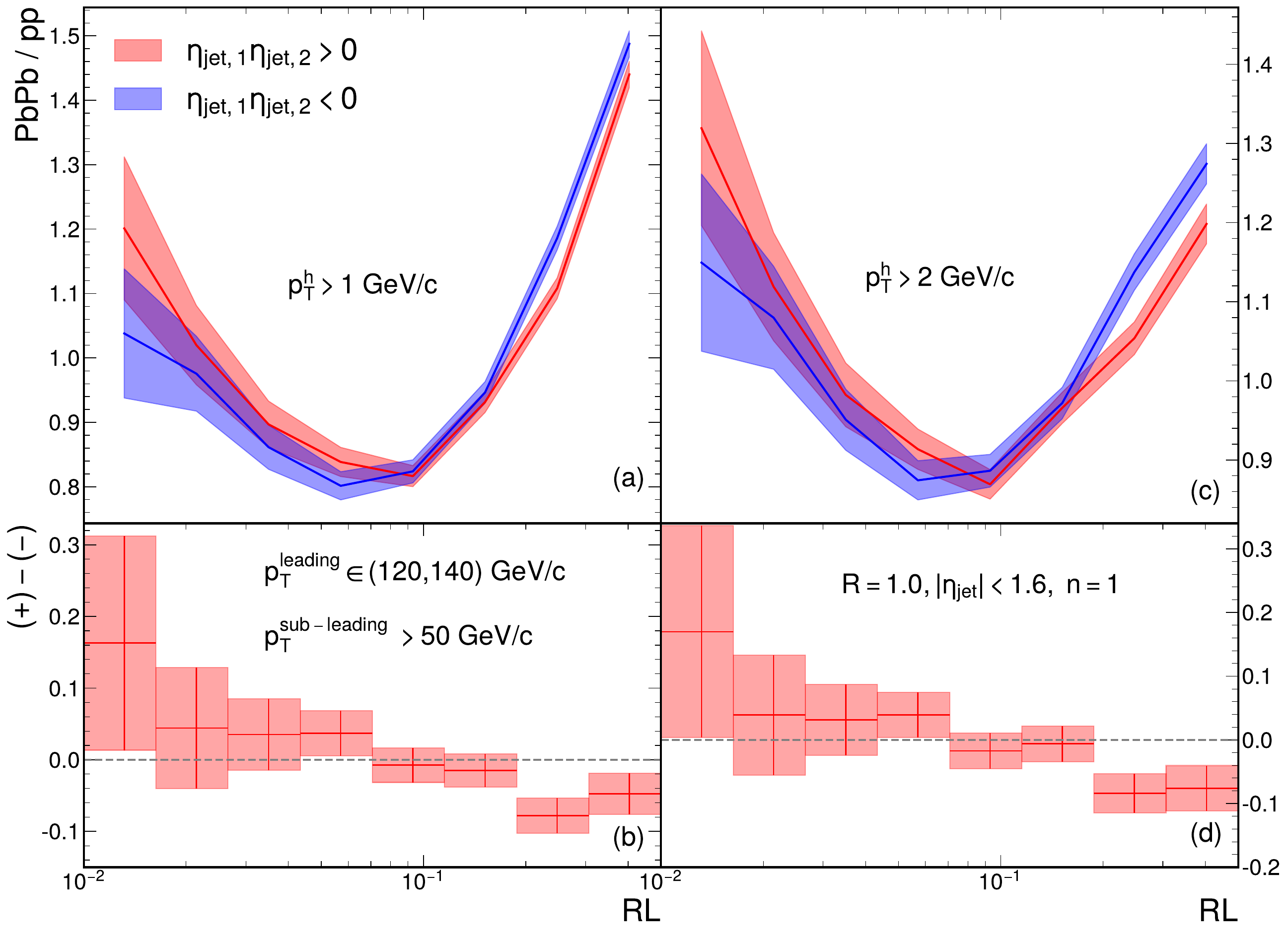}
    \caption{(a,~c) Ratio of the EEC inside leading jet in 0--10$\%$ central Pb+Pb to p+p collisions at $\sqrt{s_{\mathrm{NN}}}=5.02$~TeV for dijet events with $\eta_{\mathrm{jet},1}\eta_{\mathrm{jet},2}>0$ (red) and $\eta_{\mathrm{jet},1}\eta_{\mathrm{jet},2}<0$ (blue). The kinematic selection is $p_T^{\mathrm{jet},1}\in(120,140)$~GeV/$c$, $p_T^{\mathrm{jet},2}>50$~GeV/$c$, $|\eta_{\mathrm{jet}}|<1.6$, $R=1.0$, and $n=1$, with $p_T^h>1$~GeV/$c$ in (a) and $p_T^h>2$~GeV/$c$ in (c). (b,~d) Difference of the two ratios, $(+)-(-)$, for $p_T^h>1$~GeV/$c$ in (b) and $p_T^h>2$~GeV/$c$ in (d).}
    \label{EEC_dw}
\end{figure}

Based on this idea, we select dijet events with the kinematic cuts shown in Fig.~\ref{EEC_dw}, requiring $p_T^{\mathrm{jet},1} \in (120,140)~\mathrm{GeV}/c$, $p_T^{\mathrm{jet},2} > 50~\mathrm{GeV}/c$,  $|\eta_{\mathrm{jet}}| < 1.6$ and $|\Delta\phi_{jet,1~jet,2}|>3\pi/4$.
For jet reconstruction, the $p_T$ contribution from hydro response including both positive and negative particles is considered. In addition, the leading-jet $p_T$ interval is chosen to be narrow to enhance the sensitivity to the $p_T$ modification induced by the diffusion wake.
Meanwhile, the jet radius is increased to $R=1.0$, significantly larger than the $R=0.4$ used in the previous sections, to ensure that the diffusion wake induced by the partner jet is mostly captured within the trigger jet's cone. Since the jet $p_T$ selection bias is most pronounced in the transition region of the EEC, we restrict the analysis to $R_L < 0.5$, thereby reducing sensitivity to large-angle soft medium modifications. 
To further suppress such soft contributions, we additionally impose a track-$p_T$ cut and present results for both $p_T^{h} > 1~\mathrm{GeV}/c$ and $p_T^{h} > 2~\mathrm{GeV}/c$,
which allows us to assess the sensitivity of the result to the choice of this threshold. These selections simplify the analysis and facilitate future experimental measurements.

We then divide the selected events into two classes according to the signs of the two jet rapidities, with $\eta_{\mathrm{jet},1}\eta_{\mathrm{jet},2} > 0$ (same-side) and $\eta_{\mathrm{jet},1}\eta_{\mathrm{jet},2} < 0$ (opposite-side). The same-side class corresponds statistically to smaller dijet rapidity gaps on average, such that the two jets are close along the rapidity direction and the diffusion wake from one jet most effectively acts within the cone of the other, while the opposite-side class corresponds to a larger rapidity gap, where this effect is significantly suppressed~\cite{Yang:2025dqu}. Fig.~\ref{EEC_dw}(a,~c) shows the ratio of the EEC in Pb+Pb to p+p collisions for the two classes, and Fig.~\ref{EEC_dw}(b,~d) shows their difference, $(+) - (-)$, with $p_T^{h} > 1~\mathrm{GeV}/c$ in (a, b) and $p_T^{h} > 2~\mathrm{GeV}/c$ in (c, d). At small $R_L$, the same-side ratio lies above the opposite-side one. As $R_L$ increases, this enhancement is gradually reduced and eventually turns into a mild suppression at intermediate $R_L$. The same behavior is observed for both
$p_T^{h} > 1~\mathrm{GeV}/c$ and $p_T^{h} > 2~\mathrm{GeV}/c$. This $R_L$-dependent ordering is consistent with the effect of the jet $p_T$ selection bias in the EEC, as discussed above. The comparison between the two rapidity-gap classes therefore provides a clear manifestation of the jet-induced diffusion wake in the EEC observable.\\

\section{Summary}
In this work, we present the first CoLBT-hydro simulation of the EEC inside single-inclusive jets in 0–10$\%$ Pb+Pb collisions at $\sqrt{s_{\mathrm{NN}}}=5.02$ TeV, using an updated multi-stage parton-shower framework in which a medium scale $Q_M=2.0$ GeV separates vacuum and in-medium evolution. The calculation provides a good description of the recent CMS measurement.
By isolating the hydro-response contributions to EEC, we find that the large-angle enhancement is dominated by the modification from medium response. Implementing the CMS mixed-event background-subtraction procedure directly within the simulation, we further demonstrate that it yields results nearly identical to those obtained with the idealized theoretical subtraction, providing independent support for the robustness of the experimental method. 

To explore the path-length dependence of medium modification, we compare the EECs of leading and sub-leading jets selected according to their $p_T$ rank within each event. Sub-leading jets exhibit a systematically larger enhancement at wide angles, reflecting their stronger interaction with the QGP due to longer in-medium propagation length.

Finally, we show that the dependence of the leading-jet EEC on the dijet rapidity gap provides a direct and experimentally accessible signal of the jet-induced diffusion wake. Unlike conventional diffusion-wake observables based on soft hadrons, the signal proposed here is encoded in modifications of the jet $p_T$ and the small-angle EEC, both dominated by hard particles. This feature makes the observable experimentally cleaner and greatly facilitates future studies of jet-induced medium response and jet transport dynamics in the QGP.\\

\section{ACKNOWLEDGMENTS} 
XNW is supported in part by the National Science Foundation of China under Grant No. 12535010.
Computations in this study are performed at the NSC3/CCNU and NERSC under the award NP-ERCAP0032607. The authors also appreciate the support of the Vanderbilt ACCRE computing facility. RKE acknowledges funding from the U.S. Department of Energy, Office of Science, Office of Nuclear Physics under grant DE-SC0024660.

\bibliography{Ref}

\end{document}